\documentclass{article}
\usepackage{PRIMEarxiv}
\usepackage[utf8]{inputenc} 
\usepackage[T1]{fontenc}    
\usepackage{hyperref}       
\usepackage{url}            
\usepackage{booktabs}       
\usepackage{amsfonts}       
\usepackage{nicefrac}       
\usepackage{microtype}      
\usepackage{lipsum}
\usepackage{fancyhdr}       
\usepackage{graphicx}       
\usepackage{amsmath}
\usepackage{algorithm}
\usepackage{algpseudocode}
\usepackage{float}
\graphicspath{{media/}}     

\title{Boosting Bitcoin Minute Trend Prediction Using the Separation Index}

\author{
  \href{https://orcid.org/0009-0007-1525-6315}{\includegraphics[scale=0.06]{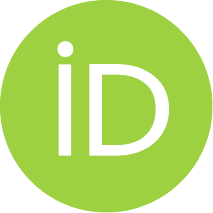}\hspace{1mm}
  Zeinab Shahsafdari} \\
  School of Electrical and Computer Engineering \\
  University of Tehran\\
  Tehran, Iran\\
  \texttt{z.shahsafdari@ut.ac.ir}\\
   \And
   \href{https://orcid.org/0000-0001-6657-6705}{\includegraphics[scale=0.06]{orcid.pdf}\hspace{1mm}
  Ahmad Kalhor} \\
  School of Electrical and Computer Engineering \\
  University of Tehran\\
  Tehran, Iran\\
  \texttt{akalhor@ut.ac.ir}\\
}

\begin{document}
\maketitle

\begin{abstract}
Predicting the trend of Bitcoin, a highly volatile cryptocurrency, remains a challenging task. Accurate forecasting holds immense potential for investors and market participants dealing with High Frequency Trading systems. The purpose of this study is to demonstrate the significance of using a systematic approach toward selecting informative observations for enhancing Bitcoin minute trend prediction. While a multitude of data collection methods exist, a crucial barrier remains: efficiently selecting the most informative data for building powerful prediction models. This study tackles this challenge head-on by introducing the Separation Index, a groundbreaking tool for fast and effective data (feature) subset selection. The Separation Index operates by measuring the improvement in class separability (i.e. upward vs. downward trends) with each added feature set. This innovative metric guides the creation of a highly informative dataset, maximizing the model's ability to differentiate between price movements. Our research demonstrates the effectiveness of this approach, achieving unprecedented accuracy in minute-scale Bitcoin trend prediction, surpassing the performance of previous studies. This significant advancement paves the way for a new era of data-driven decision-making in the dynamic world of cryptocurrency markets.
\end{abstract}

\keywords{Bitcoin Trend Prediction \and Separation Index (SI) \and High Frequency Trading (HFT) \and Technical Indicators \and Freqtrade Bot \and Lagged Data \and Deep Neural Networks (DNNs)}

\section{Introduction}
Stock market prediction, in particular, cryptocurrencies, is a classic issue and, at the same time, a challenging one for economists and computer scientists \cite{jiang2021applications}. High Frequency Trading (HFT) is an automated trading platform utilized by large investment banks, hedge funds, and institutional investors. This term emerged in the mid-2000s to refer to this type of automated trading system and has become a significant element in financial markets. It is a method of trading that uses powerful and sophisticated computer programs to execute a large number of orders in fractions of a second. HFT systems leverage sophisticated algorithms to analyze markets and identify emerging trends within fractions of a second. They can then send hundreds of stock baskets to the market at bid-ask prices that align with market changes, generating profits for traders. The decision-making process involved in selecting the best bids and offers can be highly complex and resource-intensive, depending on the specific algorithm employed. Consequently, due to this complexity, the decision-making process is not embedded in hardware but rather handled by software. Therefore, designing intelligent predictive models that can streamline the decision-making and order symbolization process is paramount in the realm of HFT \cite{leber2011high, boutros2017build}.

Over the past few decades, linear and machine learning models have been investigated with the goal of creating effective prediction models. Recently, deep learning models have been introduced as new frontiers in this field, and their development is rapidly advancing \cite{jiang2021applications}. Initially, stock market transactions are based on fundamental and technical analysis. Fundamental analysis evaluates stock prices based on their intrinsic value, while technical analysis relies solely on charts and patterns. Additionally, technical indicators derived from experience are manually crafted as input features for machine learning and deep learning models. Subsequently, linear models, including AutoRegressive Moving Average (ARMA), are introduced as solutions for stock market prediction \cite{hyndman2018forecasting}. With the development of machine learning models, techniques such as Logistic Regression (LR) and Support Vector Machines (SVM) are used for stock market prediction \cite{alpaydin2020introduction}. Deep learning, utilizing large datasets obtained from the web, parallel processing with Graphics Processing Units (GPUs), and families of neural networks such as Convolutional Neural Networks (CNNs) and Recurrent Neural Networks (RNNs), has achieved remarkable success in various applications, including time series forecasting. Deep learning models exhibit better performance than both linear and machine learning models in problems related to stock market prediction and financial time series\cite{jiang2021applications}. Recent advancements have seen the deployment of hybrid models that integrate Deep Neural Networks (DNNs) and ensemble techniques. These models are currently acknowledged as the most effective in capturing temporal relationships and achieving superior accuracy\cite{borges2020ensemble, zhou2012ensemble}.

Predicting Bitcoin trends is notoriously difficult due to the inherent characteristics of the data itself. Unlike traditional financial markets, Bitcoin operates 24/7 with a global reach, leading to a massive amount of data constantly being generated. In addition, certain data (such as minute-level data used in HFT) for some cryptocurrencies are not readily available. Collecting such data often involves scraping web resources or using APIs, which may come with challenges. Furthermore, supplementary data, including fundamental data (such as quarterly reports) and alternative data (such as company news and tweets), plays a significant role in improving modeling performance for financial data. Collecting alternative data, especially from social media platforms like Twitter, can be labor-intensive and often requires API usage, which may involve associated costs. On the other hand, feature selection from the existing pool of features using extraction and selection methods is another important challenge. In the financial domain, choosing the best feature selection and extraction algorithms and applying them to the desired dataset often involves trial and error, as it can be time-consuming.

Aside from what has been mentioned, the literature is replete with diverse methods for predicting Bitcoin's trend, each defining a variety of inputs. However, a systematic and straightforward approach for arriving at a valuable set of informative inputs for prediction is scarce. Considering the challenges of collecting a valuable dataset, this paper employs the concept of Separation Index (SI) for the first time in Bitcoin trend prediction and, without using model training, first selects a valuable set of observations from the available observations. Ultimately, it demonstrates that the dataset obtained from this method, which is gathered at a much faster speed, can be highly effective in increasing the accuracy of predictive models.

This paper is structured as follows: The Methodology section provides an introduction to the concept of SI in classification problems along with a detailed explanation of data collection and preprocessing for the Bitcoin minute dataset. Moreover, the usage of SI in subset selection and applying this method to the datasets is illustrated. Subsequently, we present experiments, results and discussion, followed by a conclusion. 

\section{Methodology}
This section delves into the application of the Separation Index (SI) for optimizing input data in classification problems. We then turn our attention to a specific case study: predicting Bitcoin minute-by-minute trends. Here, we first outline the data collection process followed by an explanation of data preprocessing. The prepared data is then evaluated using the SI metric. This metric provides a systematic and efficient way to assess how including different data subsets can enhance trend forecasting. Crucially, the SI approach fosters significant efficiency gains by reducing the necessity for iterative dataset modifications and simultaneously, leading to enhanced prediction performance.

\subsection{Using Separation Index for Observation Selection}

Separation index was first introduced in \cite{kalhor2019evaluation}. This research highlighted the efficacy and significance of the proposed index in assessing the quality of data flow within a feed-forward neural network's layered architecture. The study concluded that this straightforward, distance-based metric can be employed as an objective function for neural network classification tasks.
The Separation Index, as its name suggests, quantifies the degree of separability between data points from distinct classes.  Essentially, it assesses how effectively the features (input data points) differentiate between various class labels. Considering $Data={(x_i,l_i )}_{i=1}^{m}$ where ${x_i}$ represents each data point and may have any format such as time series, image, video, etc. and ${l_i} \in\{ 1,2,...,{n_c}\}$ where ${l_i}$ is the label of each data point and ${n_c}$ is the number of classes, this index can be defined as follows:

\begin{equation}
SI(Data) = \sum_{i=1}^{m} = \delta(l_i, l_{i^*}) 
\,\,\,\,\,\,\,\,\,\,\,\,\,\,\,\,\,\,\,\,\,\,\,\,\,\,\,\,\,\,\,
Kronecker Delta
\end{equation}

\begin{equation}
\,\,\,\,\,\,\,\,\,\,\,\,\,\,\,\,\,\,\,\,\,\,\,
i^* = \arg _{\forall q\neq i} \min ||x_i - x_q||^2
\,\,\,\,\,\,\,\,\,\,\,\,\,\,\,\,\,\,\,\,\,\,\,\,\,\,\,\,\,\,
\delta(l, l_{i^*}) = 
   \begin{cases}
            1, &         \text{if } l_i=l_{i^*}\\
            0, &         \text{if } l_i \neq l_{i^*}
    \end{cases}
\end{equation}

Note that to compute SI, ${x_i}$ must be reshaped as a vector and the input data is normalized at each dimension. $||\,\,||$ denotes Euclidean distance (L2  norm) but it may be defined as another distance. To achieve SI, the distance matrix of all data points must be computed to get the nearest neighbor for each data point. Considering the following:
\begin{equation}
\,\,\,\,\,\,\,\,\,\,\,\,
Data={(x_i,l_i )}_{i=1}^m
\,\,\,\,\,\,\,\,\,\,\,\,\,\,\,\,\,\,\,\,\,\,\,\,\,\,\,\,
x^i \in R^{n\times1}
\end{equation}
The distance matrix is calculated as below:
\begin{equation}
\,\,\,\,\,\,\,\,\,\,\,\,\,\,\,\,\,\,\,\,\,\,\,\,\,\,\,\,\,\,\,\,\,\,\,\,\,\,\,\,\,\,\,\,\,\,\,\,
D = \bigr[d_{ij}\bigr]
\,\,\,\,\,\,\,\,\,\,\,\,\,\,\,\,\,\,\,\,\,\,\,\,\,\,\,\,\,
d_{ij} = ||x_i - x_j||^2
\end{equation}
The required steps are broken down as follows:
\begin{itemize}
\item Provide data matrix: $X = \bigr[{x_1, x_2, x_3, ... ,x_m} \bigr]^T, \, X \in R^{m\times n}$
\item $M = XX^T, \, M \in R^{m\times m}$
\item $d = diag(M), \, d \in R^{m\times 1}$
\item $W = \bigr[{d, d, d, ... ,d} \bigr]^T, \, W \in R^{m\times m}$
\item Distance matrix is computed as: $D = W + W^T - 2M$
\end{itemize}
The process involves creating data and correlation matrices ($X$ and $M$), calculating the diagonal of the correlation matrix ($d$), storing diagonal values in columns in a separate matrix ($W$), and finally, using these calculations to compute a distance matrix ($D$) representing the distance between each data point.

The Separation Index is a normalized value ranging from 0 to 1.  A score of 0 indicates minimal separation between data points from different classes, while 1 signifies the maximum possible separation. This index essentially calculates the average proportion of data points where their closest neighbors share the same class label. Interestingly, this value aligns with the accuracy achievable by a nearest neighbor classifier, a simple non-parametric model. Therefore, SI serves as an informative metric that strongly correlates with the best possible accuracy a model can achieve without any data filtering beforehand.

This study investigates the application of the SI as a metric for data ranking. The objective is to identify the most valuable and informative subset of observations that are sets of features from a pool of available features, specifically tailored for prediction tasks. The initial input to the model is the entire set of observations. The final output is a reduced subset containing the most informative observations, characterized by achieving the maximum attainable SI value.  A pseudo-algorithm outlining the SI-based data ranking process is presented below.

\begin{algorithm}
\caption{SI ranking for observation selection}
\label{alg:cap}
\begin{algorithmic}
\Require Step 1

$O = \{obs_1, obs_2, obs_3, ..., obs_N \}$

$input = \{obs_{i^*} | obs_{i^*} \in O, \, max_{\arg _{i^*}}, \{ SI | \, y = prediction output\}\}$

\Require Step 2

${SI}\_{val} = \bigr[ SI(input)\bigr]$

   \>\>\>\>\>\> \texttt{\textbf{for}} {\texttt{${new}\_{obs}$} \texttt{in} \texttt{${O:}$}}
    
       \>\>\>\>\>\>\>\>\>\>\>\>   \texttt{{\textbf{if}} {{\texttt{$SI(\{input\})$} > $SI(input)$}}}
        
       \>\>\>\>\>\>\>\>\>\>\>\>\>\>\>\>\>\>\>\> {$input = \{input, {new}\_{obs}\}$
        
       \>\>\>\>\>\>\>\>\>\>\>\>\>\>\>\>\>\>\>\>{${SI}\_{val} = {SI}\_{val}{\texttt{.append}(SI(input))}$}}

\end{algorithmic}
\end{algorithm}

Following this algorithm's execution, the resulting data ranking is expected to visually resemble the diagram depicted in Figure 1. Analysis of this diagram suggests that merging observations (points in green i.e. observations 1 to 4) yields the highest achievable separability between classes and further inclusion of observations is unlikely to significantly improve prediction outcomes, and thus, omitted from the final input set (points in red).

\begin{figure}
  \centering
  \includegraphics[width=0.85\linewidth]{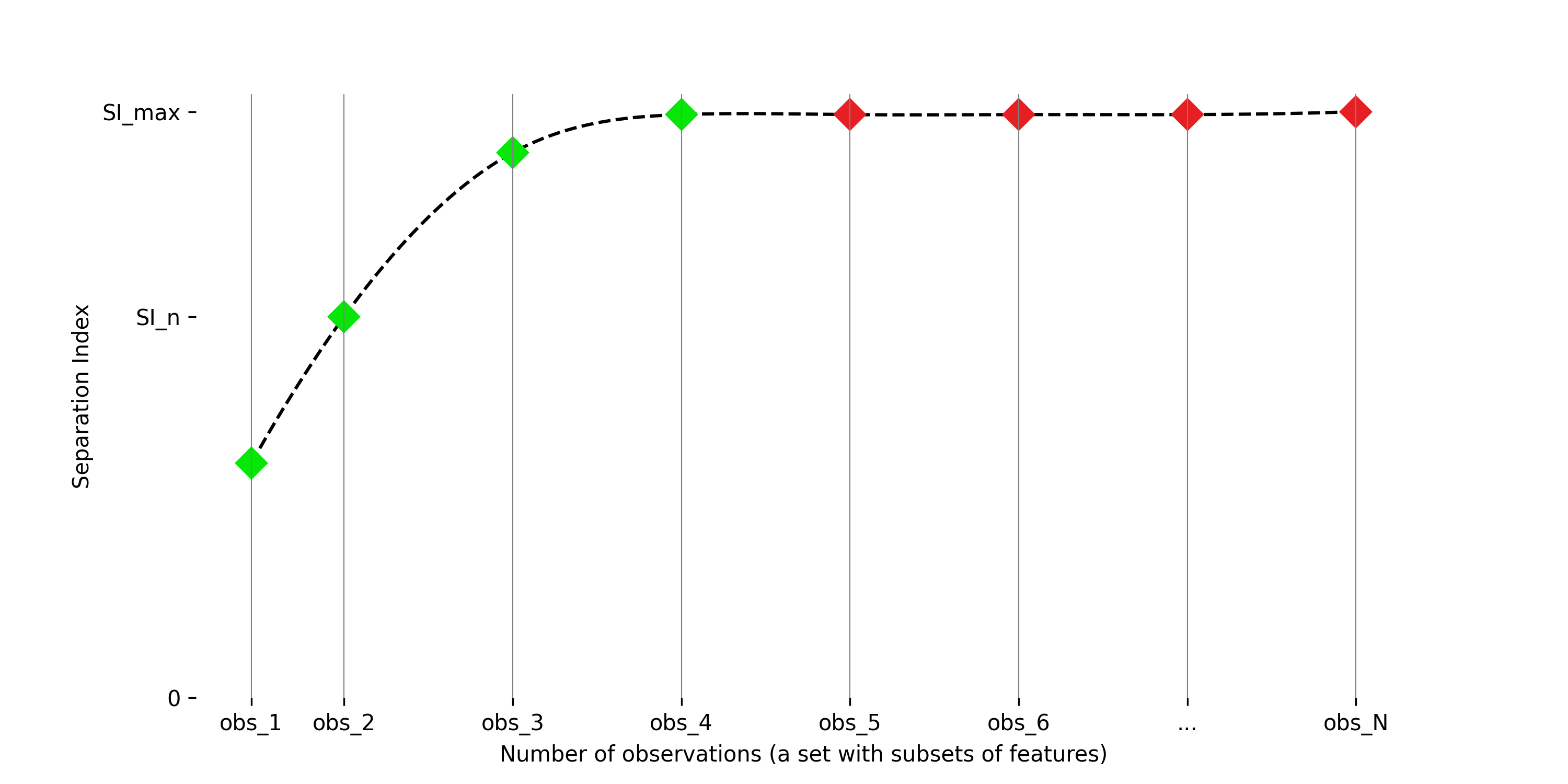}
  \caption{Expected increasing trend of selected subsets obtained by the algorithm}
  \label{fig:SI Alg}
\end{figure}

\subsection{Bitcoin Minute Data Collection and Preprocessing}
This subsection focuses on data collection and preprocessing for Bitcoin minute trend prediction. Various observations are collected, and each feature set is later evaluated using the Separation Index metric. Following data collection, the preprocessing steps involved in preparing the final input data for the prediction models are described.

\subsubsection{Data Collection}

\textbf{Freqtrade bot:} 
Freqtrade \footnote{https://www.freqtrade.io/en/stable/} is a free and publicly accessible bot for executing transactions with high frequency and implementing trading strategies for cryptocurrencies. At the outset of working with the Freqtrade framework, a comprehensive study of its documentation and related resources was conducted. To collect data using the Freqtrade tool, the KuCoin exchange has been utilized, and minute-level cryptocurrency data for Bitcoin from 2018 to 2022 has been the focus. Moreover, leveraging information available in public resources and relevant repositories on GitHub, a collection of diverse strategies designed and developed using the aforementioned framework was put to the test and evaluated. These initial steps were taken for a more precise analysis and assessment of the tool’s performance. Ultimately, from the evaluated strategies, several that appeared successful in relation to Bitcoin cryptocurrency were identified and separated for potential optimization. Among this selection, one strategy titled {BB}\_{RPB}\_{TSL} \footnote{https://github.com/jilv220/{BB}\_{RPB}\_{TSL}} was chosen as the primary and top reference strategy. This choice was based on analysis through comparison of various strategies and profitability achieved by each.

\textbf{Technical indicators with hyper-optimized parameters:} 
Greedy search and Bayesian search are two different methods for optimizing parameters in the Freqtrade tool and many other optimization tools. These methods utilize algorithms and mathematical techniques to search for the best optimal values for strategy parameters. In greedy Search, a random selection from possible parameter values is made, and the strategy is optimized based on these randomly chosen values. While this approach is simple and straightforward, it may not always lead to the ultimate optimal solution. In other words, it might find a local optimum that is not necessarily the most comprehensive. Bayesian Search, however, leverages concepts of probability and statistics to find optimal parameter values. Initially, probability distributions for parameter values are estimated based on preliminary experiences. Then, by analyzing results from back-testing and the information obtained, the probability distribution is updated, and parameter values are adjusted. This gradual approach gradually identifies the best parameter values and ultimately provides a more comprehensive optimal solution to find the global optimum for parameters. In this stage, Bayesian search is employed for the selected strategy. In continuation, a set of technical indicators existing in the selected strategy have been added to the dataset after optimizing their parameters. Table 1 shows the selected indicators. This innovative approach, employing offline parameter selection for indicators for the first time, proves crucial for efficiently identifying the optimal indicators with the most appropriate parameters from the extensive collection of available options.

\begin{table}[H]
 \caption{Selected technical indicators with hyper-optimized parameters from {BB}\_{RPB}\_{TSL}}
  \centering
  \begin{tabular}{lll}
  \toprule
    \multicolumn{2}{c}{Technical Indicators} \\
    \cmidrule(r){1-2}
    Name  &  Information     \\
    \midrule
    EMA\_8 & Exponential Moving Average - Parameter set to 8 \\
    EMA\_100 & Exponential Moving Average - Parameter set to 100 \\
    EMA\_50 & Exponential Moving Average - Parameter set to 50 \\
    EMA\_200 & Exponential Moving Average - Parameter set to 200\\
    CTI & Correlation Trend Indicator \\
    CTI\_40 & Correlation Trend Indicator - Parameter set to 40 \\
    CRSI & Connors RSI \\
    R\_96 & Williams \%R - Parameter set to 96 \\
    R\_480 & Williams \%R - Parameter set to 40 \\
    H1\_prc\_change\_5 & Range percent change or Rolling Percentage Change Maximum across interval \\
    ROC & Rate of change: $((price/prevPrice)-1) \times 100)$ \\
    RSI & Relative Strength Index \\
    CMF & Chaikin Money Flow \\
    T3 & Triple Exponential Moving Average \\
    EWO & Elliott Wave Oscillator \\
    Low\_5 & Low with shift, rolling 5 and min \\
    Safe\_dump\_50 & derived from H1\_prc\_change\_5 \\
    Weighted\_price & Volume Weighted Average Price \\
    \bottomrule
  \end{tabular}
  \label{tab:table1}
\end{table}

\textbf{Lagged data:}
Building on the research \cite{critien2022bitcoin}, this study investigated the impact of lag on minute-level trend prediction. Lag intervals of 1, 2, and 3 minutes were chosen because we aimed to predict prices one minute into the future. To create lagged datasets, we shifted the cleaned and merged data back by 1, 2, or 3 minutes respectively. While other lag values (4 and 5 minutes) were explored, the best-performing lags were 1 minute for the first model and 3 minutes for the second model (details in the next section).

\textbf{Other cryptocurrencies:}
In general, it is said that other cryptocurrencies and their fluctuations can have an impact on a particular cryptocurrency (in this case, Bitcoin). However, these impacts can vary depending on the frequency and date range, and they can be influential during one period and have no effect during another. To this end, historical data of cryptocurrencies such as Ethereum, Tether, and Litecoin were collected from the KuCoin exchange for the 4-year period (2018 to 2022) to examine their impact on Bitcoin on a minute-by-minute basis.

\subsubsection{Data Preprocessing}

\textbf{Cleaning data:}
Raw data are often lost or replaced with missing or meaningless values. To address this issue, one common approach is to remove the lost data. Another common method is to replace these missing values with statistical measures such as the mode, median, or mean, or by using interpolation techniques. The choice between these mentioned methods depends on the nature of the data, including whether it exhibits regular seasonal variations or specific trends. In this study, the second method, which is interpolation, has been utilized.

\textbf{Normalization and standardization:}
The difference between normalization and standardization lies in their respective impacts on data, particularly when dealing with outliers. Additionally, normalization is typically applied when algorithms do not have specific distribution preferences for the data, while standardization is used when the mentioned condition exists. In this study, normalization is performed with the goal of aligning dimensions across different scales in the dataset and various methods, such as StandardScaler or MinMaxScaler, are employed. In the first method, each data point is subtracted by the mean of the data and divided by the standard deviation, resulting in a mean of zero and a standard deviation of one in each dimension. In the second method, data values are scaled to a specific range (usually from zero to one) without altering the original distribution shape, achieving a consistent scaling within that particular range.

\textbf{Labeling:}
Since we're using a supervised learning approach for prediction, each data point needs a label. For this Bitcoin trend forecasting study, we added a new 'change' column to the dataset's existing features. To frame this as a binary classification problem, price increases are labeled as '1' (buy signal) and price decreases are labeled '0' (sell signal).

\section{Experiments, Results and Discussion}

\subsection{Performing SI on Collected Data}
An evaluation of how various data components contribute to Bitcoin trend prediction was conducted using the Separation Index metric on each observation set presented in Figure 2. The initial observation set comprised one-minute historical Bitcoin data obtained from the Freqtrade tool. The second set incorporated a selection of technical indicators, specifically chosen and optimized for maximum performance in predicting price movements. The third set involved a lagged dataset, created by shifting both the historical data and the technical indicator data in unison. Finally, the influence of other cryptocurrencies, including Ethereum, Tether, and Litecoin, was examined. As the Separation Index value progressively increased with the addition of each observation set, it became clear that including these carefully chosen technical indicators significantly improved the model's ability to distinguish between positive and negative price movements and trends. Merging the first three sets of observations (shown in green points), however, yielded the most acceptable Separation Index value, indicating the most effective combination for model performance. Interestingly, the analysis revealed that incorporating additional cryptocurrencies did not substantially enhance class separability. Consequently, these currencies were excluded from the final dataset (shown in red point).

This evaluation process highlights the crucial role of feature selection in model development. By meticulously analyzing the impact of each data subset using the Separation Index, valuable insights were gained into the model's anticipated performance before it was trained on the complete dataset. In essence, this approach allows for a preliminary estimation of the model's forecasting accuracy before it is deployed in a real-world setting.

\begin{figure}[H]
  \centering
  \includegraphics[width=0.95\linewidth]{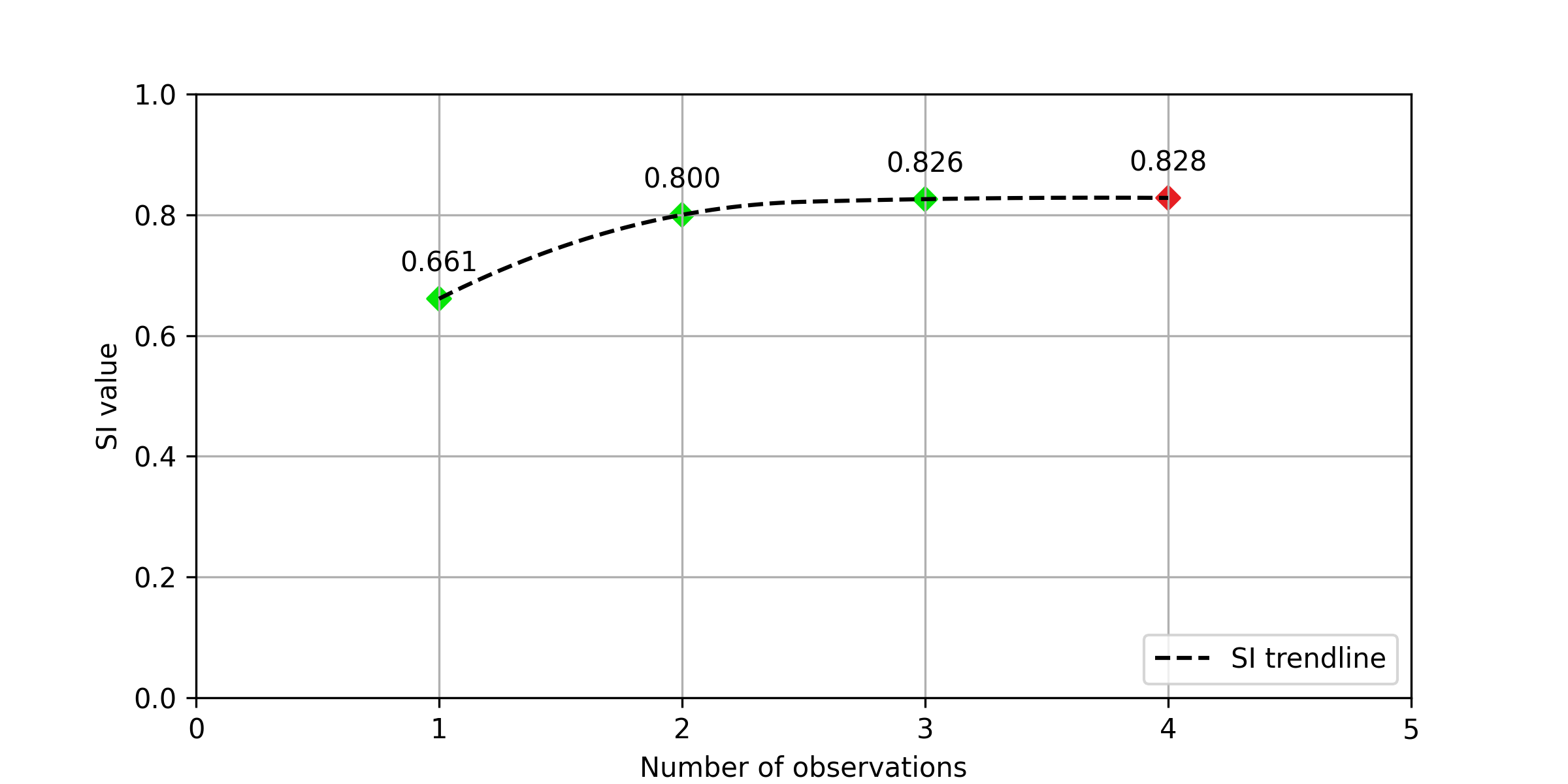}
  \caption{SI values obtained on different subsets of observations – Observation $\#$1: Bitcoin 1-minute historical data, Observation $\#$2: Technical indicators with hyper-optimized parameters, Observation $\#$3: Lagged data on all features and Observation $\#$4: Other cryptocurrencies such as Ethereum, Tether and Litecoin}
  \label{fig:SI on Bitcoin Data}
\end{figure}
It is worth mentioning that due to data availability limitations, this study couldn't incorporate fundamental data or Twitter data. Fundamental data was often only available at lower frequencies like daily updates, while Twitter data lacked coverage for the entire four-year period we examined.

\subsection{Selected Models for Bitcoin Trend Prediction}
A closer examination of previous works reveals that ensemble models provide significantly more accurate performance compared to individual models. However, the selection of a combination of different models itself is a complex issue and requires extensive testing and fine-tuning. Therefore, considering the best result of the previous works, two models have been employed: the Bidirectional Long Short Term Memory (BiLSTM) network and CNN, inspired by \cite{critien2022bitcoin}. Notably, the use of these two models and the use of a voting classifier to ensure the obtained accuracy, not only reduces the total number of parameters and overall complexity of the model but also yields higher prediction accuracy. Figure 3 depicts a flowchart featuring two models, resulting in a voting classifier that takes into account the outputs from both predictive models.
\begin{figure}[H]
  \centering
  \includegraphics[width=0.99\linewidth]{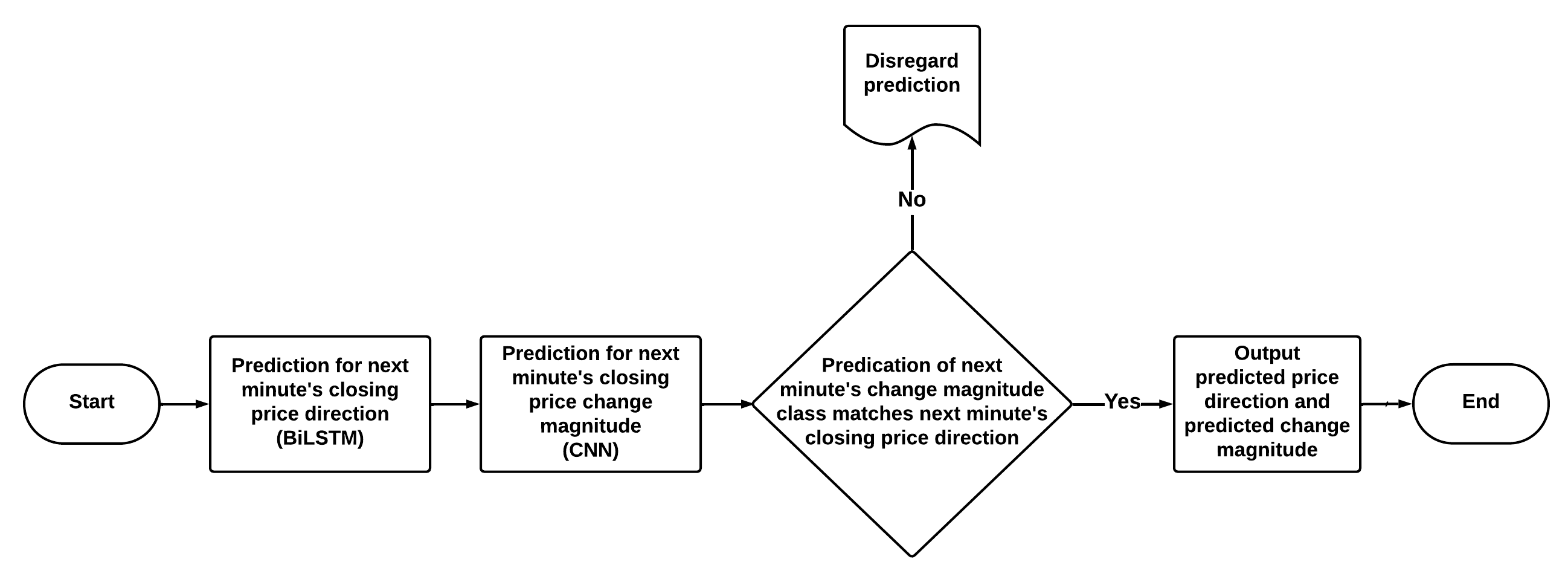}
  \caption{Minute trend prediction flowchart}
  \label{fig:flowchart}
\end{figure}
The voting classifier, as illustrated in Figure 4, operates as follows: First, it predicts the direction of the next minute’s closing price with a BiLSTM model. Next, it predicts the magnitude of the next minute’s closing price using the second model which is a CNN. Then, it checks whether the predicted direction aligns with the magnitude of the predicted change. Specifically, a match occurs if the first model outputs a 0 (indicating a price decrease), and the second model outputs a class from 0 to 4 (representing a negative magnitude of price change). Alternatively, if the first model outputs a 1 (indicating a price increase), and the second model outputs a class from 5 to 9 (representing a positive magnitude of price change), another match occurs. The prediction of the next minute’s closing price direction is retained if there is a match between the outputs of the two classifiers. 
\begin{figure}[H]
  \centering
  \includegraphics[width=0.85\linewidth]{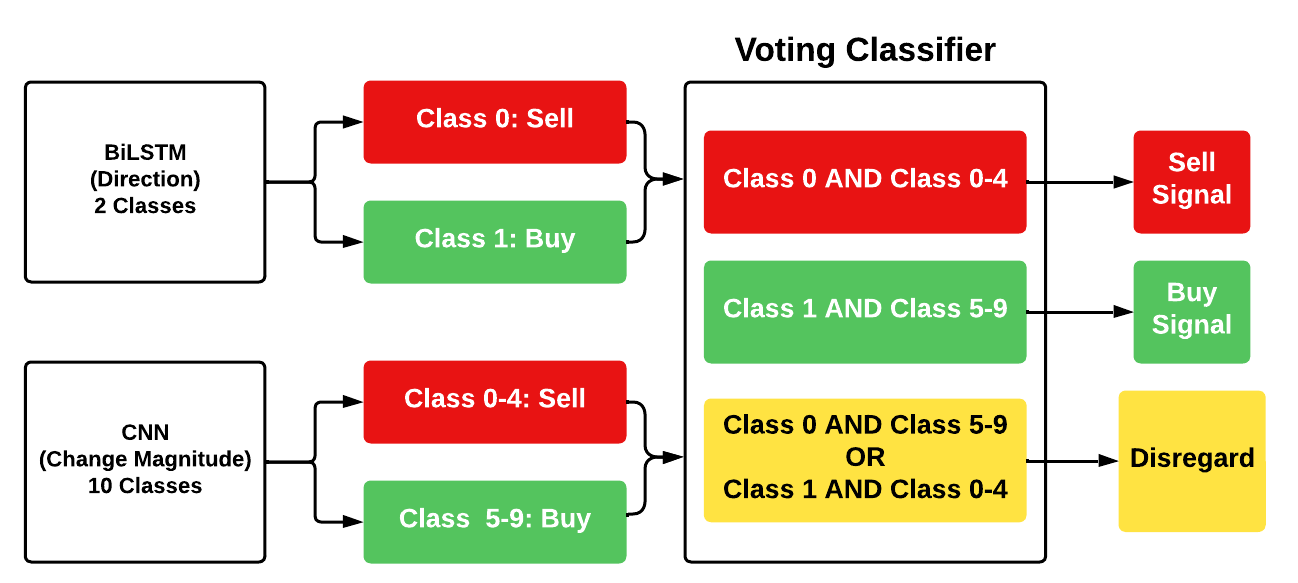}
  \caption{Voting classifier functionality}
  \label{fig:VC}
\end{figure}

\subsubsection{BiLSTM – next minute’s close price direction}
The trend or direction of the closing price can be formulated as a binary classification problem. The inputs to this network are essentially a combination of OHLCV data\footnote{A comprehensive term that includes the five critical data points — Open, High, Low, Close, and Volume — used in analyzing a financial instrument's market activity over a specified time frame} and specific technical indicators, along with lagged features (1 shift - 1 minute later) applied to all available features. Given the inputs, the prediction task is to determine whether the price will increase or decrease in the next minute.

\subsubsection{CNN – next minute’s close price change magnitude}
Another predictive model of the Convolutional Neural Network (CNN) type attempts to classify minute-level changes in closing prices as a multi-class prediction problem. This task involves predicting whether the closing price changes will fall into specific intervals. To achieve this, the minute-level closing prices are divided into different intervals using maximum and minimum price values. In other words, the mean of positive changes and the mean of negative changes in prices are calculated to define lower-class (negative changes) and upper-class (positive changes) thresholds. Additionally, classes 0 and 9 are developed to consider price changes outside these intervals (i.e., higher or lower than existing values). Table 2 presents the BiLSTM and CNN models architecture along with hyperparameter specifications and the total number of parameters used. It’s worth noting that in this study, complex models based on convolutional networks like MobileNet have been avoided due to their numerous parameters and lengthy execution times. Furthermore, these models have not been utilized in relevant research in this field so far.

\begin{table}[H]
 \caption{Hyperparameters for the best models for predicting price direction and magnitude change}
  \centering
  \begin{tabular}{lll}
    \multicolumn{3}{c}{} \\
      &  BiLSTM & CNN \\
    \midrule
    Number of layers & 2 BiLSTM & 5 Conv1D \\
    Layer size & 64 & 64 \\
    Activation Function & Linear & ReLU \\
    MaxPooling & - & 5 MaxPooling1D \\
    Pool Size & - & 2 \\
    Flatten Layer & - & 1 \\
    Dense Layer & 1 & 1 \\
    Dense Layer Units & 2 & 10 \\
    Dropout Layer & 2 & 2 \\
    Dropout Size & 0.2 & 0.2 \\
    Loss Function & Categorical Crossentropy & Categorical Crossentropy\\
    Optimization & Adam & Adam \\
    Evaluation Metric & Accuracy & Accuracy \\
    Number of Epochs & 20 & 20 \\
    Batch Size & 80 &  32\\
    Validation Split & 0.2 & 0.2 \\
    Early Stopping parameter & Validation loss & Validation loss \\
    Early Stopping patience & 5 & 5 \\
    Dataset & 1 minute lag & 1-, 2- and 3- minute lag\\
    Lagged Features & 22 & 66 \\
    Total number of Features & 44 & 88 \\
    Total number of Parameters & 143,618 & 36,554 \\
    \bottomrule
  \end{tabular}
  \label{tab:table2}
\end{table}

\subsection{Results and Comparison with Related Works}
The results of training selected models and applying them to unseen test data are showcased in Table 3. Notably, the achieved prediction accuracy closely aligns with the estimations provided by the Separation Index metric. This strong correlation validates the effectiveness of using the Separation Index for selecting informative data and ultimately improving prediction tasks.

\begin{table}[H]
 \caption{Final results for minute trend prediction}
  \centering
  \begin{tabular}{lll}
    \multicolumn{2}{c}{} \\
    Model & Accuracy(\%) on Test Data \\
    \midrule
    BiLSTM & 78.14 \\
    CNN & 79.72 \\
    Voting Classifier Max. in 15 runs & 83.70 \\
    Voting Classifier Ave. in 15 runs & 80.23 \\
    \bottomrule
  \end{tabular}
  \label{tab:table3}
\end{table}

For comparison, Table 4 presents a compilation of previous research on minute-level trend prediction alongside the findings of this study.  

\begin{table}[H]
 \caption{Related works in minute trend prediction from 2018 to 2022 along with this study}
  \centering
  \begin{tabular}{lll}
    \multicolumn{3}{c}{} \\
    \toprule
    Year and Data Range & Method(s) & Accuracy and Source \\
    
    \midrule
    2018, Jan. 2017 to Oct. 2017 &
    Neural Networks and LR &
    59.2, 58.5 and 58.2\%, \cite{saudagar2023using} \\
    
    2019, 2018 to 2019 & 
    Neural Networks &
    54.90\%, \cite{piran2023qoe} \\
    
    2019, 2013 to 2017 &
    RSM,LSTM and MLP &
    62.64, 46.88 and 47.66\%, \cite{shintate2019trend} \\
    
    2020, 2018 to 2019 &
    Hybrid CNN-LSTM &
    74.69\%, \cite{alonso2020convolution} \\
    
    2020, 2013 to 2018 &
    SVM,LR,ANN and RF &
    55-65\%, \cite{akyildirim2021prediction} \\
    
    2020, Mar. 2018 to May. 2018 &
    RF Regression &
    62.48\%, \cite{sattarov2020forecasting} \\
    
    2021, Mar. 2019 to Dec. 2019 &
    GRU,LSTM, LR, RF and Ensemble RNN &
    50.9-56\%, \cite{jaquart2021short} \\
    
    2018 to 2022 &
    A voting classifier for BiLSTM and CNN &
    80.23-83.70\%, \textbf{This Study} \\
    
    \bottomrule
  \end{tabular}
  \label{tab:table4}
\end{table}

As evidenced by the results in Table 4, leveraging a novel application of the Separation Index for observation selection, this study demonstrates a significant improvement in prediction accuracy for minute-scale Bitcoin trends. The evaluation process involved a systematic assessment of the dataset, ultimately selecting the most informative features. Subsequently, these features were incorporated into the predictive models, successfully leading to the achievement of the anticipated outcomes.

\section{Conclusion}
Existing methods for data collection have shown promise, but a key challenge lies in efficiently selecting the most impactful features for the final dataset and subsequent models. This paper addressed this gap by introducing the Separation Index as a novel metric for fast and effective data subset selection. The Separation Index measured how ranking new features improved the distinction between different classes in the final dataset. In addition, the Freqtrade tool was examined. This tool was used for extracting minute-level data and testing profitable trading strategies for Bitcoin cryptocurrency. Subsequently, a recent paper employed hybrid models such as BiLSTM, CNN, and a voting classifier for trend prediction. It achieved acceptable prediction accuracy for daily frequency data by selecting a small number of features, including historical Bitcoin data and sentiments from Twitter. Therefore, to better address the needs of a High Frequency Trading prediction (i.e., faster responsiveness), minute-level data was extracted from the Freqtrade tool. Since fundamental indicators were not available at minute-level frequency, and Twitter data was only accessible for two years, the initial focus was on Bitcoin historical data along with technical indicators. During this process, due to familiarity with the Freqtrade tool and after evaluating more than 10 different strategies for Bitcoin price prediction, the {BB}\_{RPB}\_{TSL} strategy was chosen because of its higher profitability compared to others. Furthermore, the idea was proposed to utilize the existing technical indicators in this strategy for the neural networks. However, it was crucial to use a fewer and more appropriate combination of these indicators with optimized parameters. To achieve this, after applying a Bayesian search algorithm and optimizing the parameters of the indicators in this strategy, these indicators were added to the minute-level dataset. It is noteworthy that selecting such indicators with optimized hyperparameters was done for the first time in related works. Ultimately, lagged features were added to the final dataset and the voting classifier achieved an average accuracy of 80.23\% and a maximum of 83.70\% for minute trend prediction of Bitcoin. Our study demonstrated that the Separation Index metric significantly contributed to building an informative dataset and approximately achieved the same results before training the predictive models. This, in turn, led to superior prediction results in our Bitcoin minute trend forecasting models, exceeding the performance of the previous studies. 

\vspace{0.5cm}
\textbf{Abbreviations}

ANN: Artificial Neural Network; API: Application Programming Interface; ARMA: AutoRegressive Moving Average; BiLSTM: Bidirectional Long Short-Term Memory; CNNs: Convolutional Neural Networks; DNNs: Deep Neural Networks; GPUs: Graphics Processing Units; HFT: High Frequency Trading; LR: Logistic Regression; LSTM: Long Short-Term Memory; MLP: Multi-Layer Perceptron; OHLCV: Open, High, Low, Close, and Volume; RF: Random Forest; RNNs: Recurrent Neural Networks; RSM: Response Surface Modeling; SI: Separation Index; SVM: Support Vector Machine.

\bibliographystyle{unsrt}  
\bibliography{references}  

\end{document}